\documentclass[aps,pre,twocolumn,groupedaddress,showpacs,amssymb]{revtex4}
\usepackage[dvips]{graphicx}

\newcommand{\sbr}{\rhd}
\newcommand{\xor}{\oplus}
\newcommand{\interval}[2]{\{#1:#2\}}

\begin{document}

\title{Pseudorandom number generators and the square site percolation threshold}

\author{Michael J.\ Lee}
\affiliation{Department of Physics and Astronomy, University of Canterbury, Christchurch, New Zealand}

\begin{abstract}
A select collection of pseudorandom number generators is applied to a Monte Carlo study of the two dimensional square site percolation model.
A generator suitable for high precision calculations is identified from an application specific test of randomness.
After extended computation and analysis, an ostensibly reliable value of $p_\mathrm{c} = 0.59274598(4)$ is obtained for the percolation threshold.
\end{abstract}


\maketitle

\section{Introduction}

The square site percolation threshold, $p_\mathrm{c}$, is a clearly and simply defined mathematical concept \cite{Stauffer,Bollobas}.
Percolation models have been well studied, and are known for their numerous applications \cite{Sahimi}.
Yet to date, no analytical expression has been found for the numerical value of $p_\mathrm{c}$.
The square site lattice lacks the symmetry that has allowed exact solutions on other topologies \cite{Bollobas,Harris,Kesten,Sykes,SykesJMP,Wierman,ZiffEx}.
So long as the problem remains intractable, statistical estimates from Monte Carlo studies can, at least, offer approximate values.
Such calculations invariably make extensive use of some form of pseudorandom number generator (PRNG).

A PRNG is a deterministic algorithm that outputs a sequence of words with properties closely mimicking those of a truly random sequence.
Well analysed generators include the linear congruential, lagged Fibonacci, generalised feedback shift register, and derivatives thereof \cite{Knuth,Lehmer,MarsagliaLCG,Tausworthe,Lewis,Kirkpatrick,Marsaglia,ZiffQT,Matsumoto,L'EcuyerMC,MarsagliaWEB,MarsagliaX,Tsang,Brent,L'Ecuyer}.
Because these algorithms are simple, they do not produce output with the complexity of a random sequence \cite{Kolmogorov,Chaitin}.
The autocorrelation coefficients of a pseudorandom sequence are not identically zero, and these departures from true randomness introduce a sampling bias that leads to systematic error.

Twenty years ago, concern was being given to the demands then made of PRNGs in calculations using $10^{12}$ pseudorandom numbers generated at MHz rates \cite{CompagnerJCP}.
Recently, high performance parallel computer systems with thousands, rather than tens or hundreds, of processors have become much more widely available.
These enable calculations with $10^{15}$ pseudorandom numbers generated at GHz rates, and are likely to play a central role in future research.
Very high precision can now be achieved through brute force of sampling, but accuracy is another matter.
For reliable Monte Carlo estimates at these new higher precision levels, the PRNG(s) chosen must be of sufficient quality.
Hence contemporary demands upon PRNGs are, and will continue to become, much greater than in the past.

This study compares several established PRNGs within the context of the square site percolation problem.
Following application specific testing, a seemingly reliable generator is identified.
This is subsequently used to locate the percolation threshold with, in principle, both accuracy and precision.

\section{Generators}

Throughout this study, the computational word length, $w$, shall be fixed at $32$.
All arithmetical operations taking place within any PRNG are performed in modulo $2^w$.
All PRNG arithmetical operands, and products thereof, are members of $\interval{0}{2^w-1}$, where $\interval{a}{b}$ denotes the set of all integers not less than $a$ and not greater than $b$.
Consequently the words of any PRNG output sequence also belong to $\interval{0}{2^w-1}$.
The $i$th word of an output sequence shall be denoted by $x_i$.
With one noted exception, no output sequence is decimated in any way.
The first million words of each sequence are discarded prior to beginning any Monte Carlo sampling procedure.

Some PRNGs make use of bit-wise operations within their internal algorithms.
The notation adopted here is $\xor$ for bit-wise Boolean logical exclusive-or, and $\sbr m$ for shift $m$ bits to the right (where $m$ is a positive integer).
Whenever these bit-wise operations are performed, the operands are decomposed into their respective standard binary representations, most-significant bit (leftmost) to least-significant bit (rightmost).
Arithmetic being constrained to a subset of the integers, any bits shifted to a position right of the decimal point are lost.
Hence, within this study, the operation of $\sbr m$ is equivalent to integer division by $2^m$.

The specific PRNGs considered within this exercise are defined as follows.

TT is the two-tap additive lagged Fibonacci generator $x_i = x_{i-418} + x_{i-1279}$.
This generator has previously been used for high-precision percolation threshold measurement by Newman and Ziff \cite{Newman,NewmanPRE}.

TTT combines the output from a pair of two-tap generalised feedback shift-register generators, $u_i = u_{i-471} \xor u_{i-9689}$ and $v_i = v_{i-30} \xor v_{i-127}$, to return a single word $x_i = u_i \xor v_i$.
This is the generator most likely used for two and three dimensional percolation by Deng and Bl\"ote \cite{Deng,PCBlote}.

SWB is a Marsaglia and Zaman subtract with borrow generator, $x_i = x_{i-222} - x_{i-237} - \beta_{i-1}$, where the borrow, $\beta_i$, is equal to one if $x_{i-222} < x_{i-237} + \beta_{i-1}$, and is otherwise equal to zero \cite{Marsaglia,MarsagliaWEB}.

QTA is the quad-tap generalised feedback shift-register generator $x_i = x_{i-157} \xor x_{i-314} \xor x_{i-471} \xor x_{i-9689}$, as used by Ziff and Stell \cite{ZiffLSC,Ziff} (see \cite{ZiffQT}).

QTB is the quad-tap generalised feedback shift-register generator $x_i = x_{i-471} \xor x_{i-1586} \xor x_{i-6988} \xor x_{i-9689}$.
This generator has been used by Newman and Ziff, and has been found to produce threshold estimates consistent with those of the TT generator \cite{Newman,NewmanPRE,ZiffPRE}.

XG is Brent's xorgen4096 generator \cite{Brent}.
Specifically, the implementation xorgen4096i, from his C language xorgens304 distribution, was that used here.
This generator has performed well in randomness tests conducted by L'Ecuyer and Simard \cite{L'Ecuyer}.

MT is Matsumoto and Nishimura's MT19937 Mersenne twister generator \cite{Matsumoto}.
Specifically, their MT19937ar C language distribution was the implementation used here.
The MT19937 algorithm has been used for computing integrals in semi-rigorous work by Balister, Bollob\'as, Walters and Riordan \cite{Balister,Riordan}, and for Monte Carlo sampling by Lee \cite{Lee}.

DMT is a pair of MT generators operated entirely independently of one another.
The output sequence from each of these generators is decimated, with only every fourth word used.
Lattice sites are then selected by means of their Cartesian coordinates, using one number from each generator.
This scheme has previously been used by Lee \cite{Lee}.

Let $L$ be the number of sites lying on each edge of an $L \times L$ square lattice.
Every site on that lattice is typically given a unique index or label, $j \in \interval{0}{L^2-1}$.
In a microcanonical ensemble Monte Carlo calculation \cite{Newman,NewmanPRE,ZiffPRE}, such as those performed here, PRNG output words, $x_i$, are used to pseudorandomly select sites, $s_j$, for occupation.
Now, consider a transformation $T(x,N) \equiv x \sbr (w - \log_2 N)$, where $N$ is a positive-integer power of two.
This is a distribution preserving many-to-one surjective map from the integers $x \in \interval{0}{2^w-1}$ to the integers $T(x,N) \in \interval{0}{N-1}$.
Further consider a bijection $H$ that maps the integers $\interval{0}{N-1}$ onto site labels.
With the usual choice of site labels also being the integers $\interval{0}{N-1}$, $H$ is conventionally taken to be the identity map.
In the single generator systems defined above, PRNG output word $x_i$ is associated with site $s_{j(x_i)}$ via $j(x_i) = H(T(x_i,L^2))$.
For the DMT, a pair of PRNG output words, $u_i$ and $v_i$ (one from each of the output decimated MT generators), is mapped to site $s_{j(u_i,v_i)}$ via $j(u_i,v_i) = H(T(u_i,L)+LT(v_i,L))$.
This halves the number of bits actually used from each output word (the most significant bits being those retained).

Each of these generators must be provided with an initial finite sequence of words from which to begin calculating an infinite pseudorandom sequence.
In the case of the TT generator for instance, a list of some $1279$ initial words is required.
These initial lists were constructed by one of four simpler generators, here denoted LCGa, LCGb, LCGm and WMx.
LCGa is the linear congruential generator, $x_i = 69069 x_{i-1} + 1$, suggested by Marsaglia \cite{MarsagliaLCG}.
LCGb is a similar linear congruential generator, $x_i = 69069 x_{i-1} + 1234567$, also due to Marsaglia \cite{MarsagliaWEB}.
LCGm is the modified linear congruential generator, $x_i = 1812433253(x_{i-1} \xor (x_{i-1} \sbr 30)) + i$, appearing in Matsumoto and Nishimura's MT19937ar distribution of their Mersenne twister algorithm.
WMx is the Weyl modified Marsaglia xorshift generator built into the xorgens4096i algorithm appearing within Brent's xorgens304 distribution \cite{Brent,MarsagliaX}.
These initialisation generators are themselves seeded from a single word, $x_0 \in \interval{0}{2^w-1}$.
The TTT and DMT generators both require two initialisation lists, each being derived from one of these four generators, each starting with a distinct independent seed word.

\section{Test Procedure}

The above listed generators were compared, in the context of site percolation on the square lattice, by using each one to make a Monte Carlo estimate of the crossing probability function, $R_{L,n}$, at $L = 2048$, over the domain $n \in \interval{2474000}{2498300}$.
$R_{L,n}$ is defined as the probability that a single cluster connects two specified opposing boundary sides of the $N = L \times L$ square lattice in the microcanonical ensemble when precisely $n$ random sites are occupied.
The value of $R_{2048,n}$ monotonically increases from around $0.05$ at $n = 2474000$ to around $0.95$ at $n = 2498300$.
Hence the occupation domain studied encompasses the critical region of the percolative phase transition.
The numerous lattice configurations required to accurately determine $R_{L,n}$ were constructed, from each PRNG output sequence, over the above domain only, by the unbiased algorithm of Lee \cite{Lee}.
The only exception was the XG generator from which samples were obtained over the same domain by the unbiased algorithm of Newman and Ziff \cite{Newman,NewmanPRE}.

The Newman and Ziff binomial convolution
\begin{equation}\label{ce}
R_L(p) = \sum_n {N \choose n} p^n (1-p)^{N-n} R_{L,n}
\end{equation}
then gives the crossing probability, $R_L(p)$, in the canonical ensemble where each lattice site is independently randomly occupied at probability $p$ \cite{Newman,NewmanPRE,ZiffPRE}.
In principle the summation should run over all $n \in \interval{0}{N}$, but as samples were taken only over the restricted domain of $n$ above, the summation was truncated accordingly.
The standard deviation of the binomial distribution in equation (\ref{ce}) is given by $\sigma_{L,p} \approx L\sqrt{p(1-p)}$.
The data analysis here is concerned with values of $p$ such that the distribution maximum, located at $n = \mathrm{nint}(pN)$, lies between $10$ and $12$ $\sigma_{L,p}$ from the nearest end of the sampling region.
Consequently, the truncation induced error in $R_L(p)$ is not more than $10^{-15}$.
This is completely negligible when compared to statistical sampling uncertainties, which were never less than $10^{-8}$.

The canonical crossing probability curve is used to identify a site occupation probability, $p_\mathrm{f}(L)$, defined such that
\begin{equation}\label{pe}
R_L(p_\mathrm{f}(L)) = 1/2 + k/L\mathrm{,}
\end{equation}
where $k = 0.320(1)$, as determined by Ziff and Newman \cite{ZiffPRE} (their parameter $b_0$).
For $p \approx p_\mathrm{c}$, to first order $R_L(p) \sim 0.5 + k/L + \mathrm{O}\left((p-p_\mathrm{c})L^{1/\nu}\right)$ \cite{Ziff}, and hence $p_\mathrm{f}(L)$ provides a reasonable estimate of the critical point $p_\mathrm{c}$.
Ziff and Newman have found that the second order equation $R_L(p_\mathrm{c}) \approx 0.5 + kL^{-1} - 0.44L^{-2}$ is a better model of the data at small $L$ \cite{ZiffPRE}, however the $L^{-2}$ term is negligible for $L \geq 1024$ at the levels of precision considered here.
Values for $p_\mathrm{f}(2048)$ were thus obtained from each of the PRNGs described above.
These were subsequently compared against each other and against previous $p_\mathrm{c}$ estimates made with the same generators.
For large $L$, $R_L(p)$ rises very steeply in the neighbourhood of $p \approx p_\mathrm{c}$.
Consequently, $p_\mathrm{f}(L)$ is relatively insensitive to the exact value of $k$ provided that $k/L \ll 1$.
The uncertainty in $k$ limits the maximum attainable precision in $p_\mathrm{f}(2048)$ to $\pm 2 \times 10^{-9}$.

Combinatorial terms of the binomial distribution in equation (\ref{ce}) were calculated by the essentially exact method of Newman and Ziff \cite{NewmanPRE}.
For $p$ near $p_\mathrm{c}$, use of the Gaussian approximation to the binomial would have introduced an error of order $10^{-8}$ in $R_{2048}(p)$, this corresponding to an error of order $10^{-10}$ in $p$ itself.
It is sometimes possible to dispense with the convolution altogether and make a microcanonical ensemble approximation of $R_L(p = n/N) \approx R_{L,n}$.
With $L = 2048$, and for $p$ near $p_\mathrm{c}$, this introduces an error of around $4 \times 10^{-6}$ in $R_L(p)$, which corresponds to an error of around $2 \times 10^{-8}$ in $p$.
This approximation is acceptable at low enough precision, has the advantage that only a much narrower domain of sampling need be considered, and has been employed in earlier work by Lee \cite{Lee}.
However, since the induced error (measured as the difference between $p$ and $n/N$ such that either $R_L(p) = R_{L,n} = 0.5$ or $R_L(p) = R_{L,n} = 0.5 + k/L$) was found to scale as only $L^{-1.5(3)}$, when a set of measurements are to be taken over a range of lattice sizes, to precisions of order $1/N$, the error will become significant at large $L$.
The approximation was not adopted here.

Correlations inevitably found in the output sequence of any deterministic pseudorandom number generator will result in correlations within the spatial pattern of occupied sites upon the lattice.
As noted by Compagner \cite{Compagner}, this in turn will bias the resulting Monte Carlo estimate of the crossing probability function.
Consequently, when estimates obtained from two different generators are inconsistent, then at least one of those generators likely suffers from significant correlations in its output sequence, hence rendering it unsuitable for use at the level of precision of the study.
Because the true values of $R_{L,n}$, $R_L(p)$ and $p_\mathrm{c}$ are unknown, it will be unclear as to which of the generators is deficient.

\begin{figure}
\caption{Standard (left) and (example) nonstandard (right) enumerations of the $L = 4$ square lattice.\label{ef}}
\includegraphics{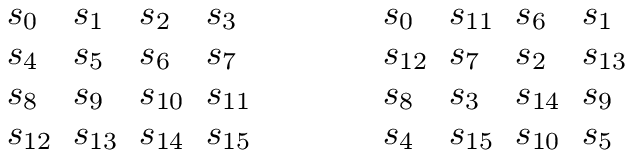}
\end{figure}

While there is merit in performing general tests on PRNGs, it is often preferable to have an application specific test such as the sensitive hull walk of Ziff \cite{ZiffQT}.
Here a scheme is used that changes the relation between numerical PRNG output sequences and the spatial patterns of occupied lattice sites, without altering the underlying problem or topology in any way.
The standard enumeration of the lattice, as shown in figure \ref{ef}, prescribes a specific relation between patterns in PRNG output and in clusters of occupied sites.
By adopting some other (nonstandard) enumeration, as per the example in figure \ref{ef}, some different relation is obtained.
An ideal random number generator will produce results independent of the chosen enumeration.
A pseudorandom number generator, with correlations in its output sequence, will produce results that do depend upon the enumeration.
By comparing results from a common generator on two different lattice enumerations, inadequate, outcome biasing, generators may be identified.
This simple application specific test does not require knowledge of the percolation threshold or spanning probability curves.

The direct approach to implementing such an enumeration is to allocate each site a set of pointers explicitly identifying its geometrical neighbours.
In the nonstandard enumeration of figure \ref{ef}, for example, $s_7$ would have pointers to $s_2$, $s_3$, $s_{11}$ and $s_{12}$.
Sites are pseudorandomly selected as per normal and the Monte Carlo sampling proceeds just as for the standard enumeration.
In practice this results in a dramatic performance decrease of the simulations (more than a factor of two was found in this study).
The problem is believed to be the cache prefetching of the high performance computer system used, where if, in some linear array, $s_j$ is being accessed then the hardware assumes that $s_{j+1}$ (being the next contiguous data element in memory) will be wanted next.

An alternative method is to change the mapping, $H$, between scaled PRNG output words, $y = T(x,N)$, and site labels, $j$.
In the standard enumeration of figure \ref{ef}, $H(y) = y$ is the identity map.
In the nonstandard enumeration, $H(0) = 0$, $H(11) = 1$, $H(6) = 2$, and so on, with the general relation being $H(y) = 3y$ (mod $16$).
This is analogous to the cluster label labels of Hoshen and Kopelman \cite{Hoshen}.
The nonstandard enumeration is that effectively in use, while the standard enumeration is preserved in computer memory, thus avoiding performance problems.

When the hash function, $H$, is simple (that is, of similar algebraic complexity to the PRNG), it will not so much hide correlations in the output sequence as manifest those patterns in some other way, giving rise to a different estimate for $p_\mathrm{f}(L)$.
If, on the (systematic) standard enumeration, correlations in PRNG output lead to spatial correlations of occupied sites that in turn bias the estimate, then, on some other (systematic) nonstandard enumeration, those same PRNG correlations will give rise to spatial correlations of a different nature that bias the estimate in some other way.
This provides a simple, application specific, test for PRNG biasing of the Monte Carlo samples.
If a given PRNG is correlation free, then the estimates derived from it will be independent of the lattice enumeration.
If instead, the PRNG output does suffer from correlations, then different enumerations may lead to different results.
The test can be tuned, with the hash chosen so as to maximise the observed shift in the test quantity.
A simple hash related to the taps or period is bound to highlight intrinsic PRNG shortcomings \cite{Compagner}.
Alternatively, a hash much more complex than the generator algorithm could go some way toward hiding output sequence correlation induced bias.

Hence define two further generators, TTH and MTH, as (respectively) the exact same TT and MT generators defined previously, but with a somewhat arbitrary non-trivial mapping $H(y) = 947y$ (mod $N$) between integers $y \in \interval{0}{N-1}$ and site labels $j \in \interval{0}{N-1}$.
On a lattice of $L = 2048$, this hash is equivalent to a systematic nonstandard enumeration where the rightward and downward neighbours of site $s_j$ are (when they exist) $s_{j+3171195 \mathrm{~}(\mathrm{mod~}N)}$ and $s_{j+1824768\mathrm{~}(\mathrm{mod~}N)}$ respectively.

\section{Test Results\label{trs}}

The various estimates of $p_\mathrm{f}(2048)$ thus obtained are listed in table \ref{rt}.
Results are separated according to the generator and initialisation scheme used.
SWBb, for instance, indicates the SWB generator with its initialisation list derived from the LCGb output sequence.
Similarly, XGx is the XG generator initialised from the WMx output sequence.
Results shown are based on surveys of order $10^8$ effectively independent samples per occupation level $n$.
Each of these sets involved the generation of order $10^{13}$ to $10^{14}$ (approaching $10^{15}$ in the xorgens case), pseudorandom numbers.

\begin{table}[tb]
\caption{Site percolation threshold estimates for the square lattice ($p_\mathrm{f}(2048)$) obtained by various pseudorandom number generators (PRNGs) as described in the text.\label{rt}}
\begin{tabular}{l@{$\qquad$}l}
\hline
\hline
PRNG  & $p_\mathrm{f}(2048)$ \\
\hline
TTa   & $0.59274627(11)$ \\
TTHa  & $0.59274588(11)$ \\
TTTab & $0.59274628(12)$ \\
SWBb  & $0.59274617(17)$ \\
QTAa  & $0.59274588(17)$ \\
QTAm  & $0.59274603(17)$ \\
QTBa  & $0.59274610(17)$ \\
QTBb  & $0.59274621(17)$ \\
XGx   & $0.59274596(15)$ \\
MTa   & $0.59274593(17)$ \\
MTm   & $0.59274585(16)$ \\
MTHm  & $0.59274598(12)$ \\
DMTmm & $0.59274597(08)$ \\
\hline
\hline
\end{tabular}
\end{table}

TTa is the TT generator initialised from LCGa.
The TTa based $p_\mathrm{f}(2048)$ estimate in table \ref{rt} is consistent with the results of Newman and Ziff that were also obtained (primarily \cite{PCZiff}) from the TT generator (see table \ref{ct}).
TTHa is the hashed TT generator, again initialised with LCGa.
The TTa and TTHa based estimates are sufficiently different to indicate the probable existence of statistically significant correlations within the TT generator output sequence.
This gives cause for concern about the use of the TT PRNG for this application at this level of precision.

\begin{table*}[tb]
\caption{Published estimates of the square site percolation threshold. The pseudorandom number generator(s) used are given where known. Generator T is a Tausworthe generator, while C is a congruential generator. TTT is the generator most likely used by Deng and Bl\"ote. References are provided for both the result and the generator whenever those come from different sources. Uncertainties are quoted as one standard deviation statistical errors, except in the semi-rigorous results of Balister, Bollob\'as and Walters ($99.99\%$ confidence bound) and of Riordan and Walters ($99.9999\%$ confidence bound). Only those results derived from currently accepted scaling relations are shown from the greater collection in Hu, Chen and Wu. This table is essentially a continuation of that appearing in Ziff and Sapoval \cite{ZiffJPA}, there going back to 1960.\label{ct}}
\begin{tabular}{l@{$\quad$}l@{$\quad$}l@{$\quad$}l@{$\quad$}l@{$\quad$}l}
\hline
\hline
Year & Ref. & Author(s) & Method & Generator(s) & Result \\
\hline
1986 & \cite{ZiffJPA} & Ziff and Sapoval & Hull-gradient & T & $0.592745(2)$ \\
1988 & \cite{ZiffLSC,ZiffQT} & Ziff and Stell & Hull-gradient & QTA & $0.5927460(5)$ \\
1989 & \cite{Yonezawa} & Yonezawa, Sakamoto and Hori & Planar crossing & & $0.5930(1)$ \\
1992 & \cite{Ziff} & Ziff & Hull-crossing & QTA & $0.5927460(5)$ \\
1994 & \cite{HuCJP} & Hu & Histogram Monte Carlo & & $0.592(8)$ \\
1995 & \cite{HuPRB} & Hu & Histogram Monte Carlo & & $0.5928(1)$ \\
1996 & \cite{Hu} & Hu, Chen and Wu & Histogram Monte Carlo & & $0.59278(2)$ \\
1996 & \cite{Hu} & Hu, Chen and Wu & Histogram Monte Carlo & & $0.59283(4)$ \\
1996 & \cite{Hu} & Hu, Chen and Wu & Histogram Monte Carlo & & $0.59267(6)$ \\
1996 & \cite{Hu} & Hu, Chen and Wu & Histogram Monte Carlo & & $0.5814(30)$ \\
1996 & \cite{Hu} & Hu, Chen and Wu & Histogram Monte Carlo & & $0.6041(30)$ \\
2000 & \cite{Newman,NewmanPRE} & Newman and Ziff & Toroidal wrapping & TT, QTB & $0.59274621(13)$ \\
2000 & \cite{Newman,NewmanPRE} & Newman and Ziff & Toroidal wrapping & TT, QTB & $0.59274636(14)$ \\
2000 & \cite{Newman,NewmanPRE} & Newman and Ziff & Toroidal wrapping & TT, QTB & $0.59274606(15)$ \\
2000 & \cite{Newman,NewmanPRE} & Newman and Ziff & Toroidal wrapping & TT, QTB & $0.59274629(20)$ \\
2000 & \cite{Newman,PCZiff} & Ziff & Hull-gradient & QTB & $0.5927465(2)$ \\
2002 & \cite{ZiffPRE} & Ziff and Newman & Planar crossing & QTB & $0.5927464(5)$ \\
2003 & \cite{Martins,PCMartins} & Martins and Plascak & Toroidal wrapping & C & $0.5927(1)$ \\
2003 & \cite{Martins,PCMartins} & Martins and Plascak & Toroidal wrapping & C & $0.5929(3)$ \\
2005 & \cite{Deng,PCBlote} & Deng and Bl\"ote & Cylindrical correlation & TTT & $0.5927465(4)$ \\
2005 & \cite{Deng,PCBlote} & Deng and Bl\"ote & Cylindrical correlation & TTT & $0.5927466(6)$ \\
2005 & \cite{Deng,PCBlote} & Deng and Bl\"ote & Cylindrical correlation & TTT & $0.5927466(8)$ \\
2005 & \cite{Deng,PCBlote} & Deng and Bl\"ote & Cylindrical correlation & TTT & $0.5927468(10)$ \\
2005 & \cite{Balister} & Balister, Bollob\'as and Walters & Semi-rigorous & MT & $0.5927(8)$ \\
2007 & \cite{Riordan} & Riordan and Walters & Semi-rigorous & MT & $0.59275(25)$ \\
2007 & \cite{Lee} & Lee & Planar crossing & MT, DMT & $0.59274603(9)$ \\
\hline
\hline
\end{tabular}
\end{table*}

TTTab is the TTT generator with its initial $u$ and $v$ lists constructed by LCGa and LCGb respectively (the two initialising generators being given two different seeds).
The TTT based estimate in table \ref{rt} is consistent with the table \ref{ct} results of Deng and Bl\"ote most likely obtained from this generator.

QTAa and QTAm are the QTA generator respectively initialised from LCGa and LCGm.
Since the QTAa and QTAm results are consistent, there is no evidence that estimates from the QTA generator are especially sensitive to initialisation.
The union of these two data sets gives an overall estimate of $p_\mathrm{f}(2048) = 0.59274595(12)$ from the QTA generator.
This value is consistent with earlier results, in table \ref{ct}, obtained by Ziff and Stell with this generator.

QTBa and QTBb are the QTB generator initialised from LCGa and LCGb respectively.
Since the QTBa and QTBb results are consistent, there is no evidence that estimates from the QTB generator are especially sensitive to initialisation.
The union of these two data sets gives an overall estimate of $p_\mathrm{f}(2048) = 0.59274616(12)$ from the QTB generator.
This is consistent with the TTa value, and also with Newman and Ziff's similar observation regarding these two generators \cite{Newman,NewmanPRE}.
Referring to table \ref{ct}, the value is also consistent with the various threshold estimates obtained by Newman and Ziff using (at least in part) this generator.

MTa is the MT generator initialised from LCGa.
MTm is the MT generator initialised from LCGm.
Since the MTa and MTm results are consistent, there is no evidence that estimates from the Mersenne twister generator are sensitive to initialisation.
The union of these two data sets gives an estimate of $p_\mathrm{f}(2048) = 0.59274589(12)$.
This is consistent with the MTHm result derived from the hashed generator in table \ref{rt}, and hence there is no evidence that correlations in the MT output sequence influence the measurement at this level of precision.
Hence the MT generator appears to be an adequate choice for the current application.
Further combining the MTHm data into the union gives an overall estimate of $p_\mathrm{f}(2048) = 0.59274593(8)$ from the MT generator.
This MT result is inconsistent with those of the TTT and (unhashed) TT generators.
The difference in results with respect to the QTB generator is no more than could be expected by chance in a data set of this size.
The SWB, QTA and XG generator based estimates are consistent with that of the MT.

DMTmm is the DMT generator with its two initial lists independently constructed, each from one of a pair of seed words, by LCGm.
The DMTmm result is consistent with the combined MT result, thereby indicating that any possible correlations between lower order bits in MT output words are insignificant at this level of precision, or at least no worse than correlations in the higher order bits.
This suggests that the single MT generator will be adequate for the purposes of this study.
Combining all four Mersenne twister based data sets; MTa, MTm, MTHm, and DMTmm, produces an estimate of $p_\mathrm{f}(2048) = 0.59274595(6)$.
This is consistent with the result of Lee, in table \ref{ct}, obtained with this same mixture of generators but in the microcanonical approximation $R_L(n/N) \approx R_{L,n}$.
The combined value does not alter any of the above conclusions regarding the consistency or otherwise of other generators with the Mersenne twister.
Regarding the previous estimate of Lee, it was observed here that $n/N$, such that $R_{2048,n} = 0.5 + k/L$ (interpolating to non-integer $n$), usually exceeds $p_\mathrm{f}(2048)$ by approximately $2 \times 10^{-8}$.
That being so, a revised estimate of the published result would be $p_\mathrm{c} = 0.5927460(1)$.
This adjustment is much smaller than statistical uncertainties.

Although the procedure used here differs from those of previous works, the results obtained are found to be consistent when the same pseudorandom number generators are used.
However, given the use of a consistent method, it has been shown that the results thus obtained can differ with the choice of generator.
The level of PRNG sensitivity will be method dependent.
The spread in results seen here is not extreme as only reasonable quality generators have been used.

The SWB, QTA, QTB, XG and MT generators are backed by strong theory \cite{Marsaglia,ZiffQT,Matsumoto,Brent} and have been extensively tested elsewhere \cite{ZiffQT,Tsang,L'Ecuyer}.
Ziff has performed a sensitive hull generating walk test upon several generalised feedback shift register generators \cite{ZiffQT}.
Two-tap generators performed poorly in this test which concluded that they best be avoided for critical applications.
Certain quad-tap generators, particularly QTB, performed very well.
Analysis indicated that QTB should outperform QTA in principle, although no obvious problems were observed in the latter.
The MT generator has passed tuned collision tests conducted by Tsang, Hui, Chow, Chong and Tso \cite{Tsang}.
The LCGa generator failed those same tests.
L'Ecuyer and Simard have recently performed thorough randomness tests upon a large assembly of PRNGs, including SWB, QTB, MT, XG and LCGa \cite{L'Ecuyer}.
The XG generator passed all tests, the MT failed in a very limited number of instances, QTB and SWB both failed a small number of times, and LCGa failed badly.
TT was not specifically tested, although two-tap generators typically performed poorly.

Results from the Mersenne twister generator have been consistent under different initialisation methods and effective lattice enumerations (hash functions).
With the observation that results from the Mersenne twister differ from those of the two-tap lagged Fibonacci generator, in the presence of evidence suggesting that the two-tap suffers from significant output correlations, and in the absence of evidence for any such correlations in the Mersenne twister output sequence, further Monte Carlo sampling within this exercise shall be performed exclusively with the MT19937 algorithm.
Note that results from the SWB, QTA, QTB and XG generators are consistent with those of the MT.

\section{Threshold Determination}

Having identified the Mersenne twister as a suitable PRNG for the problem, a more precise determination of the square site percolation threshold can now be made.
This will be based upon Monte Carlo estimates of the microcanonical $R_{L,n}$ curves for $128 \leq L \leq 4096$ (a span of some three orders of magnitude in $N$).

Data for $L \leq 1024$ was obtained exclusively from the MT generator initialised by LCGm, and Monte Carlo sampling was conducted with the algorithm of Lee \cite{Lee}.
Sampling domains were $n \in \interval{8900}{10500}$ on the $L = 128$ lattice, $n \in \interval{37300}{40400}$ on the $L = 256$ lattice, $n \in \interval{152300}{158500}$ on the $L = 512$ lattice, and $n \in \interval{615500}{627600}$ on the $L = 1024$ lattice.
The data at $L=2048$ is the combined MTa, MTm, MTHm and DMTmm data from table \ref{rt}.
As noted, that data was obtained with the same algorithm over the site occupation domain $n \in \interval{2474000}{2498300}$.
Due to hardware constraints, the $L = 4096$ data was obtained with the more memory efficient algorithm of Newman and Ziff \cite{Newman,NewmanPRE}.
For this algorithm, the entire domain, $n \in \interval{0}{N}$, is sampled, however observations were made only for $n \in \interval{9920000}{9969000}$.
Once again, the LCGm initialised MT generator was used.
Lattices of $L$ much more than $4096$ could not be accommodated by the computer system used without substantial decreases in performance.
Estimates at each $L$ are based upon between $1 \times 10^8$ (at $L=4096$) and $4 \times 10^9$ (at $L=128$) independent samples per occupation level, $n$.
These required the generation of between $10^{13}$ (at $L = 128$) and $10^{15}$ (at $L = 4096$) pseudorandom numbers.

As before, these microcanonical ensemble crossing probability curves, $R_{L,n}$, were transformed into canonical ensemble crossing probability functions, $R_L(p)$, by the convolution of equation (\ref{ce}).
Because the various microcanonical sampling domains all encompass $\pm12\sigma_{L,p}$ of the convolution region about the critical point, the domain restriction induced error in $R_L(p)$ is completely negligible for the values of $p$ considered here.

Several statistics were calculated from each $R_L(p)$ curve.
These were Ziff's median-$p$ critical point estimator \cite{Ziff}, $p_\mathrm{m}(L)$, defined such that
\begin{equation}\label{pme}
R_L(p_\mathrm{m}(L)) = 1/2\mathrm{,}
\end{equation}
the Reynolds, Stanley and Klein real-space renormalisation group cell-to-cell estimator \cite{ReynoldsJPA,Reynolds}, $p_\mathrm{cc}(L)$, defined such that
\begin{equation}\label{pcce}
R_L(p_\mathrm{cc}(L)) = R_{L/2}(p_\mathrm{cc}(L))\mathrm{,}
\end{equation}
the Ziff and Newman linear combination estimator \cite{ZiffPRE}, $p_\mathrm{h}(L)$, defined as
\begin{equation}\label{phe}
p_\mathrm{h}(L) \equiv (p_\mathrm{m}(L) + \alpha p_\mathrm{cc}(L))/(1 + \alpha)\mathrm{,}
\end{equation}
and the real-space renormalisation group cell-to-site fixed point estimator of Reynolds, Klein and Stanley \cite{ReynoldsJPC}, $p_\mathrm{r}(L)$, defined such that
\begin{equation}\label{pfre}
R_L(p_\mathrm{r}(L)) = p_\mathrm{r}(L) \mathrm{.}
\end{equation}
Numerical estimates for these quantities are shown in table \ref{st}.

The estimators $p_\mathrm{m}$ and $p_\mathrm{cc}$ are believed to approach their limiting values on the infinite lattice as $L^{-1-1/\nu}$, where $\nu = 4/3$ \cite{Ziff,ZiffPRE}.
The estimator $p_\mathrm{h}$ is believed to converge to its limit at a faster rate of $L^{-1-\omega-1/\nu}$, where Ziff and Newman have determined a value of $\omega = 0.90(2)$ \cite{ZiffPRE} for the scaling exponent proposed by Aharony and Hovi \cite{Aharony,Hovi}.
The estimator $p_\mathrm{r}$ is believed to approach its limit as $L^{-1/\nu}$, a much slower rate of convergence than for the other estimators \cite{ZiffPRE}.
Although each of these four limits is numerically equivalent to the percolation threshold $p_\mathrm{c}$, it will be useful to adopt a general notation indicating the origin of any threshold estimates.

\begin{table*}[tb]
\caption{Site percolation threshold estimators on square lattices of various sizes $L$. $p_\mathrm{m}(L)$ is the median-$p$ estimator, $p_\mathrm{cc}(L)$ is the cell-to-cell estimator, $p_\mathrm{h}(L)$ is the linear combination estimator, and $p_\mathrm{r}(L)$ is the fixed point estimator. Results were obtained with the Mersenne twister pseudorandom number generator.\label{st}}
\begin{tabular}{l@{$\quad$}l@{$\quad$}l@{$\quad$}l@{$\quad$}l@{$\quad$}l}
\hline
\hline
$L$ & $p_\mathrm{m}(L)$ & $p_\mathrm{cc}(L)$ & $p_\mathrm{h}(L)$ & $p_\mathrm{r}(L)$ \\
\hline
128  & $0.59266108(21)$ &                &                & $0.59598352(23)$ \\
256  & $0.59272062(18)$ & $0.5928085(4)$ & $0.5927460(5)$ & $0.59467466(18)$ \\
512  & $0.59273860(15)$ & $0.5927651(4)$ & $0.5927462(4)$ & $0.59389258(15)$ \\
1024 & $0.59274377(14)$ & $0.5927514(4)$ & $0.5927460(4)$ & $0.59342699(15)$ \\
2048 & $0.59274528(06)$ & $0.5927475(2)$ & $0.5927459(2)$ & $0.59315051(06)$ \\
4096 & $0.59274573(10)$ & $0.5927464(3)$ & $0.5927459(3)$ & $0.59298626(10)$ \\
\hline
\hline
\end{tabular}
\end{table*}

\begin{figure}[tb]
\caption{Parametrised fit of first order scaling theory (equation (\ref{mse}) with $b = 0$) to experimental data ($p_\mathrm{m}(L)$ of table \ref{st}) for the median-$p$ critical point estimator.\label{sf}}
\includegraphics[width=65mm]{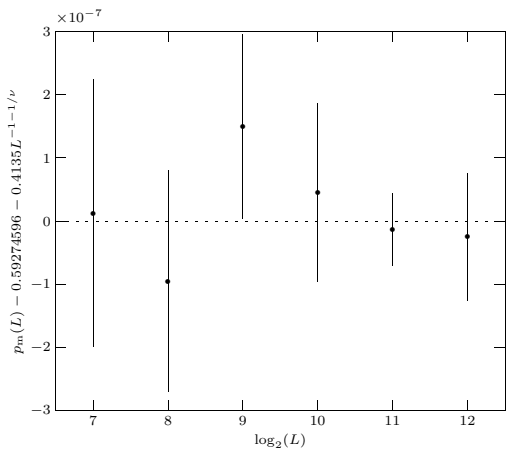}
\end{figure}

To second order, the finite size scaling relation for the median-$p$ estimator is
\begin{equation}\label{mse}
p_{\mathrm{m}}(L) \approx p_\mathrm{m}^*-aL^{-1-1/\nu}+bL^{-1-\omega-1/\nu}\mathrm{.}
\end{equation}
A parametrised fit of equation (\ref{mse}) to the data of table \ref{st} produces $p_\mathrm{m}^* = 0.59274595(4)$, $a = 0.413(5)$, and $b = 0.0(4)$.
That the coefficient $b$ is indistinguishable from zero suggests that a first order model (equation (\ref{mse}) with $b$ constrained to zero) is appropriate for the data.
As such, a precise value for $\omega$ is unimportant.
In this first order case, the coefficients are evaluated as $p_\mathrm{m}^* = 0.59274596(3)$, and $a = 0.4135(7)$.
The very good agreement between model and experiment is shown in figure \ref{sf}.
An empirical power law fit of the form $p_{\mathrm{m}}(L) \sim p_\mathrm{m}^*-aL^z$ yields $p_\mathrm{m}^* = 0.59274595(5)$, $a = 0.42(2)$, and $z = -1.75(8)$.
That the value of $z$ is indistinguishable from the assumed exponent of $-1-1/\nu$, further supports scaling of the form $L^{-1-1/\nu}$ as being the appropriate model for the data at this level of precision.
All three $p_\mathrm{m}^*$ estimates are in good agreement with one another.

\begin{figure}[tb]
\caption{Parametrised fit of first order scaling theory (equation (\ref{cse}) with $c = 0$) to experimental data ($p_\mathrm{cc}(L)$ of table \ref{st}) for the cell-to-cell critical point estimator.\label{cf}}
\includegraphics[width=65mm]{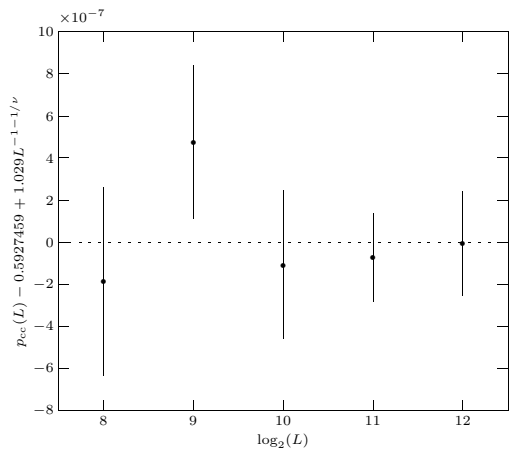}
\end{figure}

The second order scaling relation for the cell-to-cell estimator is given by
\begin{equation}\label{cse}
p_{\mathrm{cc}}(L) \approx p_\mathrm{cc}^*+\frac{a}{\alpha}L^{-1-1/\nu}+cL^{-1-\omega-1/\nu}
\end{equation}
where $\alpha \equiv 1-2^{-1/\nu}$ \cite{ZiffPRE}.
The quality of the cell-to-cell data is lower than that of the median-$p$ data, as each point is obtained from the intercept of two lines, with statistical uncertainties, at a shallow angle, and as $p_\mathrm{cc}(L)$ and $p_\mathrm{cc}(L/2)$ are not entirely independent.
A parametrised fit of equation (\ref{cse}) to the data of table \ref{st} produces $p_\mathrm{cc}^* = 0.5927458(2)$, $a = 0.441(5)$ and $c = -9(8)$.
Coefficient $c$ is not inconsistent with zero, and a first order model (equation (\ref{cse}) with $c$ constrained to zero) does fit the data, as shown in figure \ref{cf}, with coefficients of $p_\mathrm{cc}^* = 0.5927459(2)$ and $a = 0.417(4)$, in good agreement with the median-$p$ estimator results.
An empirical power law fit of the form $p_{\mathrm{cc}}(L) \sim p_\mathrm{cc}^*-(a/\alpha)L^z$ yields $p_\mathrm{cc}^* = 0.5927458(2)$, $a = 0.34(8)$, and $z = -1.71(4)$, consistent with the assumed $L^{-1-1/\nu}$ scaling relation.
All three $p_\mathrm{cc}^*$ estimates are consistent with each other and with the estimates for $p_\mathrm{m}^*$, although the precision is significantly lower.

\begin{figure}[tb]
\caption{Parametrised fit of scaling theory (equation (\ref{hse})) to experimental data ($p_\mathrm{h}(L)$ of table \ref{st}) for the linear combination critical point estimator.\label{hf}}
\includegraphics[width=65mm]{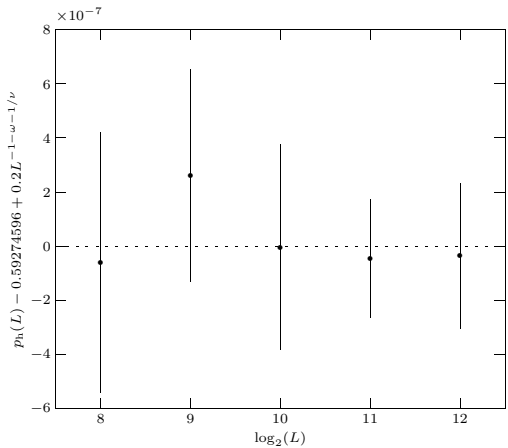}
\end{figure}

The linear combination estimator of equation (\ref{phe}) was constructed by Ziff and Newman \cite{ZiffPRE} so as to cancel the first order terms of equations (\ref{pme}) and (\ref{pcce}), leaving a faster approach to the percolation threshold;
\begin{equation}\label{hse}
p_\mathrm{h}(L) \sim p_\mathrm{h}^* + \frac{b + \alpha c}{1 + \alpha}L^{-1-\omega-1/\nu}
\end{equation}
(to first order).
A parametrised fit of this expression to the data of table \ref{st} is shown in figure \ref{hf} and produces $p_\mathrm{h}^* = 0.59274596(7)$, and $(b + \alpha c) = 0.3(9)$.
The threshold result is in good agreement with those obtained from the median-$p$ estimator data.
The value of $(b + \alpha c)$ is also in agreement, although this is not saying much given the large uncertainties.
That this value is essentially indistinguishable from zero is a reflection of the rapid rate of convergence of the $p_\mathrm{h}(L)$ estimator with $L$, as suggested by equation (\ref{hse}), and the relative lack of precision in the $p_\mathrm{h}(L)$ data.
This is unsurprising given that no higher order terms were apparent in either the $p_\mathrm{m}(L)$ or $p_\mathrm{cc}(L)$ data sets.
As such, the data was inadequate to empirically test the assumed scaling exponent and is even consistent with $p_\mathrm{h}(L) = \textrm{constant} = p_\mathrm{h}^*$, for which fitting the weighted mean gives $p_\mathrm{h}^* = 0.5927460(1)$.

\begin{figure}[tb]
\caption{Parametrised fit of scaling theory (equation (\ref{rse})) to experimental data ($p_\mathrm{r}(L)$ of table \ref{st}) for the renormalisation group fixed point percolation threshold estimator.\label{rf}}
\includegraphics[width=65mm]{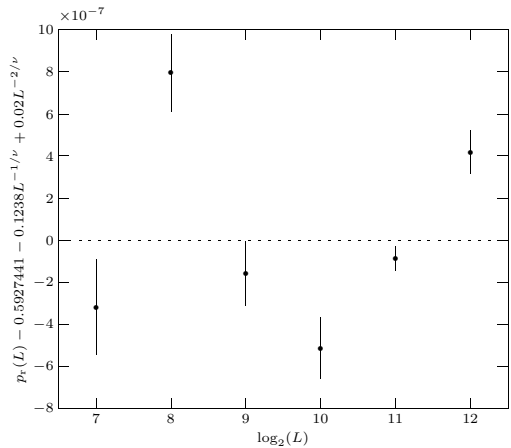}
\end{figure}

The second order scaling relation for the fixed point renormalisation group estimator is given by
\begin{equation}\label{rse}
p_\mathrm{r}(L) \sim p_\mathrm{r}^* + rL^{-1/\nu} + sL^{-2/\nu}
\end{equation}
\cite{ZiffPRE}, and so has a slower rate of convergence to its infinite lattice limit than any of the other estimators above.
A fit to the data of table \ref{st} yields well defined numerical values for the coefficients; $p_\mathrm{r}^* = 0.5927441(7)$, $r = 0.1238(2)$, and $s = -0.021(6)$.
However, as shown in figure \ref{rf}, the model of equation (\ref{rse}) is but a loose match to the data at best, with higher order terms evidently remaining significant.
As such, the stated uncertainty in $p_\mathrm{r}^*$ is misleading and will be addressed shortly within the next section.
An empirical power law fit of the form $p_{\mathrm{r}}(L) \sim p_\mathrm{r}^*-rL^z$ yields $p_\mathrm{r}^* = 0.592743(2)$, $r = 0.1219(8)$, and $z = -0.748(2)$, consistent with the assumed first order exponent of $-1/\nu$.
The first order model (equation (\ref{hse}) with $s$ constrained to zero) returns $p_\mathrm{r}^* = 0.5927456(8)$ and $r = 0.1233(1)$.
Of course, neither of these two functions describe the data any better than does the second order model.

\section{Robustness}

Several estimates have now been made for the square site percolation threshold, $p_\mathrm{c}$, using all the data of table \ref{st} and with varying degrees of precision.
Of these, the most precise is $p_\mathrm{c} = p_\mathrm{m}^* = 0.59274596(3)$, obtained from the median-$p$ estimator data using the first order scaling model $p_\mathrm{m}(L) = p_\mathrm{m}^* + aL^{-1-1/\nu}$.
It is prudent to establish the robustness of the results with respect to variations in the data and in the assumed model, over what domains the various models are valid, and how the domain and any fixed model parameters influence the estimate of $p_\mathrm{c}$.
There is a trade off between fitting to as much data over as great a domain as possible, so as to reduce statistical sampling fluctuations and hence to refine the result, and fitting to only data from large lattices where finite size effects are smaller and the scaling theories better describe the data.
The results in table \ref{rt} are consistent, where they overlap at $L = 128$ and $L = 256$, with those of Ziff and Newman \cite{ZiffPRE}.
Hence their data was used to extend the domain down to $L = 8$ as necessary.

The fixed point renormalisation group estimator is possessed of good quality data, but has a slow rate of convergence to its limit $p_\mathrm{r}^*$.
The $p_\mathrm{r}(L)$ data of table \ref{st} can be reasonably well fit with the addition of an $L^{-3/\nu}$ term to the model, however the coefficient of $L^{-4/\nu}$, in an even higher order model, is not zero.
Values of the coefficients fluctuate with the order of the model, suggesting that even higher order terms remain significant.
Empirical power law fits are consistent with the leading order exponent being $-1/\nu$, however a purely first order model does not fit the data well until the domain is truncated to $L \geq 256$.
Results for $p_\mathrm{r}^*$ are sensitive to the presence or absence of individual data points, the $L = 4096$ point altering the result by $\pm 1 \times 10^{-6}$.
Under different models and data ranges, threshold estimates range from $0.592744$ to $0.592746$.
The difference is much larger than the uncertainty in the individual estimates and so not much  weight should be given to those.
Consequently, although the raw data at a given $L$ is relatively precise, the slow rate of convergence of the fixed point renormalisation group estimator leads to only a very rough figure of $p_\mathrm{r}^* = 0.592745(1)$.

The linear combination estimator suffers from relatively large statistical uncertainties in the data, and points are not entirely independent of one another.
However, the estimator does claim a very rapid rate of convergence to its limit, $p_\mathrm{h}^*$.
The model of equation (\ref{hse}) fits the data well for $L \geq 32$.
Results thus obtained range from $p_\mathrm{h}^* = 0.59274594(5)$ to $p_\mathrm{h}^* = 0.59274603(8)$, with the presence or absence of individual data points making differences of as much as $\pm 4 \times 10^{-8}$ in $p_\mathrm{h}^*$.
Allowing for alternative values of the parameter $\omega$, between $0.85$ and $0.95$, the estimate changes by no more than $\pm 1 \times 10^{-8}$.
The data is not precise enough to either support or falsify the assumed scaling relation, and is not inconsistent with $p_\mathrm{h}(L) = \mathrm{constant}$.
Even so, all estimates for $p_\mathrm{h}^*$ were consistent with one another and with the $128 \leq L \leq 4096$ $p_\mathrm{h}(L)$ data mean of $0.5927460(1)$.
Hence the linear combination estimator appears to be robust, and the mean value, which covers the entire range of results, should be a more than safe estimate for $p_\mathrm{h}^*$.
The value of $p_\mathrm{h}^* = 0.59274596(7)$, obtained from all the $p_\mathrm{h}(L)$ data of this study, should be reliable.

With similarly low data quality, non-independent points, and a slower rate of convergence, the cell-to-cell renormalisation group estimator should not be expected to provide any refinement in $p_\mathrm{c}$ over the linear combination approach.
Over domains where the various models fit the data, cell-to-cell results for $p_\mathrm{cc}^*$ range from $0.5927458(2)$ to $0.5927461(2)$.
Sensitivity to the presence or absence of individual data points is as for the linear combination results, but here this is much smaller than statistical uncertainties.
The estimate $p_\mathrm{cc}^* = 0.5927459(2)$, obtained earlier from fitting the first order scaling model to all the $p_\mathrm{cc}$ data of table \ref{st}, is in agreement with the entire range of cell-to-cell results above, and so is robust, if imprecise.

The median-$p$ based estimates have the same rate of convergence as the cell-to-cell estimates, but with independent data points of much higher quality.
The median-$p$ estimates are less sensitive to the presence or absence of any one particular data point, this making a difference of at most $2 \times 10^{-8}$, and typically of less than $1 \times 10^{-8}$, in the result for $p_\mathrm{m}^*$.
The first order model fits the data for $L \geq 128$, with results lying in the range $p_\mathrm{m}^* = 0.59274594(3)$ to $p_\mathrm{m}^* = 0.59274596(4)$.
The second order model fits the data for $L \geq 16$, with results lying between $p_\mathrm{m}^* = 0.59274591(8)$ and $p_\mathrm{m}^* = 0.59274600(5)$.
The empirical power law model makes a good fit for $L \geq 64$, with estimates of $p_\mathrm{m}^*$ running from $0.59274589(2)$ up to $0.59274603(2)$, and scaling exponents in the range $-1.729(7)$ to $-1.79(2)$.
As noted in the previous section, the first order fit matches the data of table \ref{st} very well, the empirical fit agrees with the assumed exponent of $-1-1/\nu$, and coefficients of higher order terms were insignificant.
This indicates that the first order model does indeed provide an accurate description for the finite-size scaling behaviour of the median-$p$ estimator.
The estimate thus obtained, of $p_\mathrm{m}^* = 0.59274596(3)$, does not quite encompass the entire range of results above.
Allowing for an extreme scenario, where even the model and scaling exponent may not be quite right, a more conservative figure of $p_\mathrm{m}^* = 0.59274596(4)$ does cover all of the above results.
Hence this final value of the median-$p$ estimate for $p_\mathrm{c}$ should be quite dependable.
Incidentally, a standard error of $4 \times 10^{-8}$ in $p_\mathrm{c}$ is approximately what would be expected from the total amount of data sampled in this study (as listed in the $p_\mathrm{m}(L)$ column of table \ref{st}).
Parameter $a$ of equation (\ref{mse}) shows much more sensitivity to the data domain and model than does $p_\mathrm{m}^*$.
The fitted value given in the previous section was the most precise obtained.
An overall result of $a = 0.415(5)$ is more reasonable in light of the other estimates.

The four estimators have now produced equally many robust estimates for the two-dimensional square site percolation threshold $p_\mathrm{c}$.
As summarised in table \ref{ert}, these are $p_\mathrm{r}^* = 0.592745(1)$, $p_\mathrm{cc}^* = 0.5927459(2)$, $p_\mathrm{h}^* = 0.59274596(7)$, and $p_\mathrm{m}^* = 0.59274596(4)$, in good mutual agreement.
Taking $p_\mathrm{c} = 0.59274596$, and returning to the canonical spanning probability curves, a good match between the data of $128 \leq L \leq 4096$ and the theory of $R_L(p_\mathrm{c}) \approx 0.5 + kL^{-1} + \mathrm{O}(L^{-2})$ was had for $k = 0.317(1)$.
No higher order terms were seen, with the coefficient of $L^{-2}$ being indistinguishable from zero.
The value of $k$ found here is a little lower than those of Ziff, $k=0.319(1)$ \cite{Ziff}, and Newman and Ziff, $k=0.320(1)$ \cite{NewmanPRE}.
The difference in $p_\mathrm{f}(2048)$ resulting from using $k = 0.317$, as opposed to $k = 0.320$, in equation (\ref{pe}) is around $6 \times 10^{-9}$.
This is much less than the statistical uncertainties in the results of table \ref{rt}, upholding the claimed insensitivity of those estimates to $k$.
Hence those results remain reasonable (PRNG biased) estimates for $p_\mathrm{c}$, and the direct comparison with earlier $p_\mathrm{c}$ estimates is valid.
The estimate $a = 0.415(5)$ found above is consistent with the results of Ziff, Newman, Hovi and Aharony \cite{Ziff,ZiffPRE,Hovi} ($a$ here equates to their ratio $b_0/a_1$).
Since $k$ equates to $b_0$, it follows that $a_1 = 0.76(1)$ from the data obtained within this exercise.
This estimate is also consistent with those of previous works \cite{Ziff,ZiffPRE,Hovi}.

The above results are based on data acquired solely from the Mersenne twister, that generator having been determined as suitable for this problem.
In section \ref{trs}, results obtained from the SWB, QTA, QTB and XG generators were found to be consistent with results obtained from the Mersenne twister.
Although those four generators were not tested to the same extent as Mersenne twister, there is no objective reason to discount them entirely.
Incorporating the data obtained from these generators earlier leads to revised values of $p_\mathrm{m}(2048) = 0.59274532(4)$, $p_\mathrm{r}(2048) = 0.59315055(4)$, $p_\mathrm{cc}(2048) = 0.5927476(1)$, $p_\mathrm{cc}(4096) = 0.5927463(2)$, $p_\mathrm{h}(2048) = 0.5927459(1)$, and $p_\mathrm{h}(4096) = 0.5927459(3)$ for the various estimators of table \ref{st}.
Note that the majority of the data remains Mersenne twister based.

Use of these revised values does not alter either the fixed point limit, $p_\mathrm{r}^*$, or the cell-to-cell limit, $p_\mathrm{cc}^*$.
The linear combination limit is raised to $p_\mathrm{h}^* = 0.59274598(6)$, an adjustment of rather less than its statistical uncertainty.

\begin{figure}[tb]
\caption{Parametrised fit of empirical power law model $p_\mathrm{m}(L) = p_\mathrm{m}^* - aL^z$ to combined experimental data from the Mersenne twister, subtract with borrow, xorgens, and both quad-tap generators.\label{zf}}
\includegraphics[width=65mm]{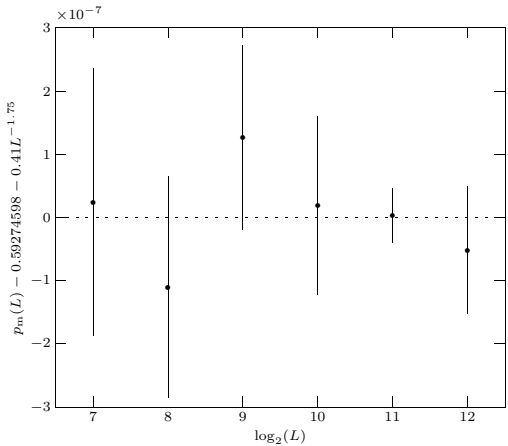}
\end{figure}

A parametrised fit of equation (\ref{mse}) to the revised median-$p$ data yields $p_\mathrm{m}^* = 0.59274598(3)$, $a = 0.415(7)$, and $b = 0.1(5)$.
As before, the coefficient of the higher order term is indistinguishable from zero.
A first order fit (of equation (\ref{mse}) with $b$ constrained to zero) produces $p_\mathrm{m}^* = 0.59274598(3)$, and $a = 0.414(1)$.
An empirical power law fit of the form $p_{\mathrm{m}}(L) \sim p_\mathrm{m}^*-aL^z$ finds $p_\mathrm{m}^* = 0.59274598(4)$, $a = 0.41(2)$, and $z = -1.75(1)$.
The excellent agreement between this model and the experimental data is shown in figure \ref{zf}.
The fitted value of $z$ is indistinguishable from the assumed scaling exponent of $-1-1/\nu$ (with $\nu = 4/3$).
All fitted parameters are consistent across the three models.

Performing robustness checks as before, the first order model fits the data for $L \geq 128$, with results lying within a worst case range of $p_\mathrm{m}^* = 0.59274598(6)$ to $p_\mathrm{m}^* = 0.59274599(8)$, and much more typically within $p_\mathrm{m}^* = 0.59274598(2)$ to $p_\mathrm{m}^* = 0.59274599(4)$.
The second order model fits the data for $L \geq 16$, with results lying between $p_\mathrm{m}^* = 0.59274596(3)$ and $p_\mathrm{m}^* = 0.59274602(4)$.
The empirical power law model makes a good fit for $L \geq 128$, with estimates of $p_\mathrm{m}^*$ running from $0.59274598(4)$ up to $0.59274599(8)$, and scaling exponents, $z$, in the range $-1.74(1)$ to $-1.77(1)$.
Hence the data supports the validity of the first order model with the assumed scaling exponent, and a standard error of $3 \times 10^{-8}$ in $p_\mathrm{m}^*$ appears fully justified.
The various threshold estimates are summarised in table \ref{ert}.
Using the revised data, and $p_\mathrm{c} = 0.59274598$, the estimate of the finite size correction parameter remains unchanged at $k = 0.317(1)$.
Nor is any significant change is seen in parameter $a$.

\begin{table}[tb]
\caption{Infinite lattice limit estimates for the percolation threshold. Results are shown, by estimator, for the Mersenne twister only data (MT, MTH, DMT), and also for the combined generators data (MT, MTH, DMT, SWB, QTA, QTB, XG).\label{ert}}
\begin{tabular}{l@{$\quad$}l@{$\quad$}l}
\hline
\hline
Limit & Mersenne & Combined \\
\hline
$p_\mathrm{r}^*$  & $0.592745(1)$ & $0.592745(1)$ \\
$p_\mathrm{cc}^*$ & $0.5927459(2)$ & $0.5927459(2)$ \\
$p_\mathrm{h}^*$  & $0.59274596(7)$ & $0.59274598(6)$ \\
$p_\mathrm{m}^*$  & $0.59274596(4)$ & $0.59274598(3)$ \\
\hline
\hline
\end{tabular}
\end{table}

Assuming the suitability of the Mersenne twister PRNG for this particular Monte Carlo application, and also assuming that the median-$p$ estimator approaches the critical point as $p_\mathrm{m}(L) - p_\mathrm{c} \propto L^{-1-1/\nu}$, where $\nu = 4/3$, as supported by the data, then a robust estimate for the square site percolation threshold is $p_\mathrm{c} = 0.59274596(4)$.
A value for $\omega$ is not required.
Further assuming the suitability of the subtract with borrow, xorgens and both quad-tap generators, the additional data adjusts this estimate to $p_\mathrm{c} = 0.59274598(3)$.
Continuing to assume the reliability of those generators, while dropping the assumed scaling exponent and requiring only that $p_\mathrm{m}(L) - p_\mathrm{c} \propto L^z$, for some $z$, the estimate becomes $p_\mathrm{c} = 0.59274598(4)$.
These three estimates are mutually consistent to well within statistical uncertainties.
The most precise of them has a standard error of $3 \times 10^{-8}$, however a degree of caution is warranted in that none of the generators were tested to that level of precision.
That being the case, this study's final estimate for the square site percolation threshold is
\begin{equation}
p_\mathrm{c} = 0.59274598(4)\mathrm{.}
\end{equation}

This, primarily Mersenne twister based, estimate is consistent with almost all previous results in table \ref{ct}.
In particular, it is in good agreement with the Mersenne twister derived estimate of Lee.
Taken collectively however, those results, excluding that of Lee, would suggest a higher value for $p_\mathrm{c}$, in the vicinity of $0.5927463(1)$.
Although the value obtained here lies well outside of that range, the difference could be attributable to the various pseudorandom number generators used.
While it is not impossible that the result obtained here may reflect some detectable influence of the chosen generators, precautions against this were taken and no evidence of bias was found.

\section{Conclusions}

Increasing availability of highly parallel computer facilities now makes it practical to obtain significant quantities of Monte Carlo data from large lattices.
This allows for greater precision in derived statistics, but requires very good quality pseudorandom number generators as it is well established that inadequate generators lead to erroneous results.

Tests were performed upon several generators and it was found that use of simple two-tap generators should probably be avoided for this application.
The MT19937 generator appeared to be suitable and was adopted for the majority of the Monte Carlo sampling conducted within this study.
No dependence was found upon the (reasonable) choice of generator initialisation.

Percolation threshold estimates subsequently made from various crossing probability statistics were found to be in good mutual agreement.
The most precise of these was obtained from the median-$p$ estimator.
Data quality was such that precise results could be obtained without the need to assume a particular scaling exponent.
Even so, results were in good agreement with a leading exponent of $-1-1/\nu$ and no higher order term was found.
The square site percolation threshold was subsequently determined to be $p_\mathrm{c} = 0.59274598(4)$.

This estimate is consistent with the majority of earlier results on an individual basis, but not with those same results combined.
Evidence suggests, however, that at least some of those earlier results have been influenced by the pseudorandom number generators used.
The generators used here appear to be of adequate quality, and the main generator, MT19937, passed an application specific test of randomness.
Furthermore, efforts were made to ensure the reliability of the error bounds in that final estimate, which should, then, be accurate.

\section{Acknowledgements}

The author would like to thank
R.\ M.\ Ziff for raising the need to address the percolation problem with different generators, for discussions of the results and for comments on the manuscript.
Also H.\ W.\ J.\ Bl\"ote, R.\ P.\ Brent and P.\ H.\ L.\ Martins for providing details on their respective pseudorandom number generators, and A.\ J.\ E.\ Dale, G.\ L.\ Evans and C.\ J.\ McMurtrie for assistance with BlueFern, the University of Canterbury's Blue Gene supercomputing facility, upon which the Monte Carlo sampling was conducted.

\end{document}